\documentstyle[aps,twocolumn,prl]{revtex}

\begin{document}

\input{epsf}

\draft

\title{Tunnelings as Catastrophes}
\author{C.\ A.\ A.\ de Carvalho\cite{CAC}}
\address{Instituto de F\'\i sica, Universidade Federal do Rio de Janeiro,
\\ Cx.\ Postal 68528, CEP 21945-970, Rio de Janeiro, RJ, Brasil}
\author{R.\ M.\ Cavalcanti\cite{RMC}}
\address{Departamento de F\'\i sica, Pontif\'\i cia Universidade
Cat\'olica do Rio de Janeiro, \\ Cx.\ Postal 38071, CEP 22452-970,
Rio de Janeiro, RJ, Brasil}
\maketitle

\begin{abstract}

We use path-integrals to derive a general expression for the
semiclassical approximation to the partition function of a
one-dimensional quantum-mechanical system. Our expression depends
solely on ordinary integrals which involve the potential. For high
temperatures, the semiclassical expression is dominated by single
closed paths. As we lower the temperature, new closed paths
appear, including tunneling paths. The transition from single to
multiple-path regime corresponds to well-defined catastrophes.
Tunneling sets in whenever they occur. (Our formula fully accounts
for this feature.)

\end{abstract}


We may use path-integrals in the description of Quantum Statistical
Mechanics.The partition function for a one-dimensional
quantum-mechanical system, for instance, can be expressed as
\begin{mathletters}
\begin{eqnarray}
& &Z(\beta)=\int_{-\infty}^{\infty}dx_0
\int_{x(0)=x_0}^{x(\beta\hbar)=x_0}[Dx(\tau)]\,
{\rm e}^{-S/\hbar}\;, \label{Z} \\
& &S=\int_{0}^{\beta\hbar}d\tau\,
\left[\frac{1}{2}M\dot{x}^2+V(x)\right]\;.
\end{eqnarray}
\end{mathletters}
For an arbitrary potential, $V(x)$, this integral has been
approximated by means of perturbative\cite{Feynman-Hibbs,Feynman},
variational\cite{Feynman-Hibbs,Feynman} and numerical\cite{Creutz}
techniques. Here, we shall concentrate on the semiclassical
approximation to the integral in order to: i) derive a general
formula for $Z_{\rm sc}(\beta)$ that does not require a detailed
knowledge of the classical\cite{Classical} motion;
ii) discuss the onset of tunneling
and relate it to the study of singularities in Catastrophe Theory.

The semiclassical evaluation of (\ref{Z}) yields:
\begin{equation}
\label{Zsc}
Z_{\rm sc}(\beta)=\int_{-\infty}^{\infty}dx_0
\sum_{j=1}^{N(x_0,\beta)}{\rm e}^{-S_j(x_0,\beta)/\hbar}
\Delta_j^{-1/2}(x_0,\beta)\;,
\end{equation}
where $S_j(x_0,\beta)$ denotes the action and $\Delta_j(x_0,\beta)$
represents the determinant of the fluctuation operator,
\begin{equation}
\label{F}
\hat{F}=-M\frac{d^2}{d\tau^2}+V''[x_j(\tau)]\;,
\end{equation}
both calculated at the $j$-th classical trajectory, $x_j(\tau)$,
satisfying the boundary conditions, $x_j(0)=x_j(\beta\hbar)=x_0$;
$N(x_0,\beta)$ is the number of classical trajectories which are
minima of the action functional.

The action has a simple expression in terms of the turning points
of the classical trajectory:
\begin{eqnarray}
\label{S}
S_j(x_0,\beta)&=&\beta\hbar\,V(x_{\pm}^j)\pm
2\int_{x_0}^{x_{\pm}^j}dx\,Mv(x,x_{\pm}^j) \nonumber \\
&+&2n\int_{x_-^j}^{x_+^j}dx\,Mv(x,x_{\pm}^j)\;.
\end{eqnarray}
Here, $v(x,y)\equiv\sqrt{(2/M)[V(x)-V(y)]}$ and
$x_+^j$ $(x_-^j)$ are turning points to the right (left) of $x_0$.
The first term in (\ref{S})
corresponds to the high-temperature limit of $Z(\beta)$,
where the classical paths collapse to a point, i.e.,
$x_{\pm}^j\rightarrow x_0$. For motion in regions where the
inverted potential is unbounded (hereafter called unbounded motion),
$n=0$. For periodic motion, $n$ counts the number of periods and the
second term is absent. For bounded aperiodic motion, all terms
are present. The last two terms will be negligible for potentials
which vary little over a thermal wavelength,
$\lambda=\hbar\sqrt{\beta/M}$. However, by decreasing the temperature,
quantum effects become important.

For trajectories having a single turning point $(n=0)$,
$x_{\pm}^j$ are given implicitly by
\begin{equation}
\label{t}
\tau(x_0,x_{\pm}^j)\equiv 2\int_{x_0}^{x_{\pm}^j}\frac{dx}
{v(x,x_{\pm}^j)}=\pm\beta\hbar\;,
\end{equation}
and the fluctuation determinant by
\begin{equation}
\label{Delta}
\Delta_j(x_0,\beta)=
\pm\frac{4\pi\hbar[V(x_{\pm}^j)-V(x_0)]}
{MV'(x_{\pm}^j)}\,\left[
\frac{\partial\tau(x_0,y)}{\partial y}\right]_{y=x_{\pm}^j}\;.
\end{equation}
The formula above is valid only for trajectories with a single
turning point. However, this is not really a restriction, since
trajectories with two or more turning points are naturally
excluded from our calculations, as it will become clear in the
sequel.

Formulae (\ref{S}) and (\ref{Delta}) do not require full
knowledge of the classical trajectory, as the dependence on $x_j(\tau)$
comes only through the turning point. To see how this can
simplify the evaluation of $Z_{\rm sc}(\beta)$, let us take
the harmonic oscillator, $V(x)=\frac{1}{2}M\omega^2 x^2$, as an
example. In this case, given $x_0$ and $\beta$ there is only one
trajectory, with a single turning point, given by
$x_+(x_-)=x_0/\cosh(\beta\hbar\omega/2)$ for $x_0<0\;(>0)$.
$S_1$ and $\Delta_1$
can also be readily calculated, the final result being
\begin{eqnarray}
\label{ZHO}
Z_{\rm sc}(\beta)&=&\int_{-\infty}^{\infty}dx_0\,
{\rm e}^{-(M\omega x_0^2/\hbar)\tanh(\beta\hbar\omega/2)} \nonumber \\
& &\qquad\times\;\sqrt{\frac{M\omega}{2\pi\hbar\sinh(\beta\hbar\omega)}}\;,
\end{eqnarray}
which in this case is exact.

A more interesting situation occurs in the case of the anharmonic
oscillator, $V(x)=\lambda(x^2-a^2)^2$. For $x^2>a^2$ only
single paths with single turning points exist for fixed $x_0$
and $\beta$. However, there is also a region,
$x^2<a^2$, where the classical motion is bounded and a much richer
structure exists, one in which more
than one classical path may exist for given
values of $x_0$ and $\beta$.

In a region of bounded classical motion
(a well in $-V$), such as $x^2<a^2$ for the anharmonic oscillator,
the number of classical trajectories changes as the temperature drops.
If $0\le\beta<\pi/\hbar\omega_m$ (where
$\omega_m\equiv\sqrt{-V''(x_m)/M}$ and $x_m$ is a local minimum
of $-V$), for every $x_0$ in this region
there is only one closed path, with a single turning point,
satisfying the classical equations of motion.
For $x_0<x_m\;(>x_m)$
this path goes to the left (right) and returns to $x_0$.
For $x_0=x_m$, it sits still
at the bottom of the well. It is this single-path
regime which goes smoothly into the high-temperature limit.

For $\beta=\pi/\hbar\omega_m$, the solution that sits still
at $x_m$ becomes unstable. Its fluctuation operator is that of
a harmonic oscillator with $\omega^2=\omega_m^2$. Its finite
temperature determinant is $\Delta(x_m,\beta)=
2\pi\hbar\sin(\beta\hbar\omega_m)/M\omega_m$.
This goes through zero at $\beta=\pi/\hbar\omega_m$ and becomes negative for
$\beta>\pi/\hbar\omega_m$, thus signaling that
$x(\tau)=x_m$ becomes unstable. At the same time, two new classical paths
appear, symmetric with respect to $x_m$, as depicted in Fig.\ \ref{fig1}.
Therefore, at $x_0=x_m$, we go from a
single-path regime to a triple-path regime as we cross
$\beta=\pi/\hbar\omega_m$. The two new paths are degenerate minima, whereas
the path that sits still at $x_m$ becomes a saddle-point of the
action, with a single negative mode.

As $\beta$ grows beyond $\pi/\hbar\omega_m$, an analogous situation
occurs for other values of $x_0$ inside the well. When the fluctuation
determinant around the classical path for a given $x_0$ vanishes,
a new classical path appears. Its single turning point lies opposite,
with respect to $x_m$, to that of the formerly unique path. It may
be interpreted as a tunneling trajectory, since it traverses a
classically forbidden region of $V$.
If $\beta$ is increased further, this tunneling
trajectory splits into two, as illustrated in Fig.\ \ref{fig2}.
One is a local minimum of the action, whereas the other is a
saddle-point, with only one negative mode. Again, we have
transitioned from a single to a triple-path regime. As $\beta$ grows,
the triple-path region spreads out around
$x_m$. The frontiers of that region are defined by the
points $x_0$ such that
$\left[{\partial\tau(x_0,y)}/{\partial y}\right]_{y=x_{\pm}^j}=0$.

The phenomenon just described is an example of catastrophe.
It takes place whenever the number of classical trajectories
changes, with one or more of them becoming unstable, as we lower
the temperature. Conversely, we may say that the phenomenon is
characterized by the coalescence of two or more of the classical
trajectories, as we increase the temperature. This is akin to
the ocurrence of caustics in Optics\cite{Berry}, where light
rays play the role of classical trajectories and the action
is replaced with the optical distance.

If we denote by $s_1$ the coordinate associated with the direction
of instability of our example, the action can be viewed as
$S=S(s_1,\ldots;x_0,\beta)$. If we use the basis of eigenfunctions
of the fluctuation operator around classical paths, $s_1$ will
correspond to the direction along the one eigenfunction whose
eigenvalue goes through zero, with the dots referring to all others.
Catastrophe Theory\cite{Berry,Thom,Saunders} allows us to write
down the ``normal form'', $S_N$,
of this generating function in the three-dimensional subspace
made up by the unstable direction of state
space (the set of paths) and the two control variables (related to
$\beta$ and $x_0$); it is this subspace which
is relevant for the study of the onset of instability. As catastrophes
are classified by their codimension in control space, which here is
two-dimensional, only those of codimension 1 (the fold) or 2 (the cusp)
can occur. The pattern of extrema then leads to the cusp,
whose generating function is
\begin{equation}
\label{SN}
S_N(s_1;u_1,v_1)=s_1^4+u_1s_1^2+v_1s_1\;,
\end{equation}
where $u_1$ and $v_1$, the control parameters, are related to
$\beta$ and $x_0$, respectively. This catastrophe is defined by
\begin{equation}
\label{SN'}
\frac{\partial S_N}{\partial s_1}=
\frac{\partial^2S_N}{\partial s_1^2}=0\;.
\end{equation}
Eliminating $s_1$ from these equations, we can draw the bifurcation
set in control space, as well as the pattern of extrema of $S_N$
(Fig.\ \ref{fig3}).

We can also plot the classical action (i.e., the action for classical
trajectories) as a function of $x_0$ for different values of
$\beta$, by exploiting the relation between the cusp and
the swallowtail catastrophes. The latter has a generating function
whose normal form is
\begin{equation}
\label{W}
W(s_1;a,b,c)=s_1^5+as_1^3+bs_1^2+cs_1\;;
\end{equation}
$a$, $b$, $c$ are control variables. The extremum condition
is, then, ${\partial W}/{\partial s_1}=0$.
The identifications $S_N\equiv -c/5$, $u_1\equiv 3a$ and
$v_1\equiv 2b$, will make the additional condition that defines
the swallowtail, $\partial^2W/\partial s_1^2=0$, coincide with
the requirement $\partial S_N/\partial s_1=0$. In the usual jargon, the
bifurcation set for the swallowtail coincides
with the equilibrium hypersurface of the cusp.
We can, then, draw the graphs for $S_N$ versus $v_1$, for various
values of $u_1$, shown in Fig.\ \ref{fig4}(a,b). Note that $v_1$ is related to
$(x_0-x_m)$ and $u_1$ to $(\beta-\pi/\hbar\omega_m)$; for small
values of both, we expect the relations to be linear.

New tunnelings, accompanied by new
catastrophes, will occur as we keep increasing $\beta$. From
$\beta=2\pi/\hbar\omega_m$, we start having tunneling trajectories
with $\dot{x}(0)=0$, i.e., with one full period. For these, the
determinant of fluctuations vanishes because the first, not the
second bracket of (\ref{Delta}) goes trough zero. However, the fluctuation
operator around such tunneling trajectories already has a negative
eigenmode. This follows from the fact that the zero-mode,
given by $\dot{x}_{\rm cl}(\tau)$, has a node. Thus, it
cannot be the ground-state for the associated Schr\"odinger
problem. This means that a new catastrophe
takes place along a second direction in function space.

This new catastrophe is also a cusp. At $x_m$,
as $\beta$ becomes larger than $2\pi/\hbar\omega_m$,
the solution that sits still acquires a second negative eigenvalue
--- thus becoming unstable along a new direction in function space ---
and two other solutions appear. They both have a negative
eigenvalue along the first direction of instability
and a positive eigenvalue along the new direction of instability.
If we combine the information from both directions,
we find that we go from two minima and a saddle-point with only
one negative eigenvalue (hereafter, a one-saddle) to two minima,
two one-saddles and one two-saddle. The two one-saddles are
degenerate in action and represent time-reversed periodic paths. As
$\beta$ increases beyond $2\pi/\hbar\omega_m$, the same phenomenon
takes place for $x_0$ around $x_m$: the one-saddles that already
existed for $\beta<2\pi/\hbar\omega_m$ in the three-path
region around $x_m$ become unstable along a second direction,
and a five-path region grows around $x_m$, with two minima,
two (time-reversed, degenerate in action) one-saddles and a
two-saddle.

The new cusp can be cast into normal form using the second direction of
instability:
\begin{equation}
\label{SN2}
S_N(s_2;u_2,v_2=0)=s_2^4+u_2s_2^2\sim(s_2^2+\frac{u_2}{2})^2\;.
\end{equation}
The absence of the linear term in Eq.\ (\ref{SN2}) [compare with
Eq.\ (\ref{SN})] reflects the degeneracy in action of the two
one-saddles. However, there is another degeneracy: since $V$ does
not depend on $\beta$, if $x_{\rm cl}(\tau)$ is a solution of the
Euler-Lagrange equations, so is $x_{\rm cl}(\tau+\tau_0)$. If
$x_{\rm cl}(\tau)$ is periodic, $x_{\rm cl}(\tau+\tau_0)$
describes the same path --- and so has the same action ---, but with
another starting point.
This can be represented in Eq.\ (\ref{SN2}) by
the choice $u_2\propto [(x_0-x_m)^2+k(2\pi/\hbar\omega_m-\beta)]$,
with $k>0$. See Fig.\ \ref{fig4}(c).

As we approach $\beta=3\pi/\hbar\omega_m$, the situation becomes
similar to the one near $\beta=\pi/\hbar\omega_m$. The difference is
that we now have to deal with a tunneling path with more
than one and less than two periods. Near $\beta=4\pi/\hbar\omega_m$,
two-period tunnelings intervene, a situation similar to that at
$\beta=2\pi/\hbar\omega_m$. The pattern which develops is depicted
in Fig.\ \ref{fig5}.

Despite the fact that we keep adding new extrema as we lower the
temperature, many of which represent tunneling solutions, only
two of these are stable (i.e., minima). In a semiclassical
approximation in euclidean time, these are the only extrema we
have to sum over, meaning that $N(x_0,\beta)$ never exceeds two.
The local minimum is a tunneling solution, whereas the global one
becomes the unique solution in the high-temperature regime.
For $\beta<\pi/\hbar\omega_m$, there will be regions with either one
or two minima of the classical action. The transition from a
single-minimum to a double-minimum region occurs at values of $x_0$
where the fluctuation determinant vanishes due to the appearance of
a caustic, thus leading to a singularity in
the integrand of Eq.\ (\ref{Zsc}). This is not a disaster,
however, as this singularity is integrable.
(Such a singularity is an
artifact of the semiclassical approximation.
It disappears in a more refined approximation\cite{Berry,DV,Weiss},
in which one includes cubic and quartic fluctuations in the direction of
function space where the instability sets in.)
Explicit calculations for a number of potencials, and the derivation
of Eq.\ (\ref{Delta}), will be presented in a future publication.

We thank Roberto Iengo, for useful conversations, Ildeu de Castro Moreira,
for pointing out Ref.\cite{Saunders}, and Flavia Maximo, for help with
the figures. Part of this work was carried out at the ICTP, and had
financial support from CNPq.

\begin{figure}
\caption{(a) Stable paths at $x_m$ for $\beta<\pi/\hbar\omega_m$ (1)
and $\beta>\pi/\hbar\omega_m$ (2 and 3); (b) Sketch of how the
extrema change along the unstable direction in function space.}
\label{fig1}
\end{figure}

\begin{figure}
\caption{(a) Stable paths at $x_m$ for $\beta<\beta_{\rm c}$ (1),
$\beta=\beta_{\rm c}$ (1 and 2\&3) and $\beta>\beta_{\rm c}$
(1,2 and 3), $\pi/\hbar\omega_m<\beta_{\rm c}<2\pi/\hbar\omega_m$;
(b) Sketch of how the extrema change along the unstable direction
in function space.}
\label{fig2}
\end{figure}

\begin{figure}
\caption{Bifurcation set for the cusp; pattern of extrema shown
schematically.}
\label{fig3}
\end{figure}

\begin{figure}
\caption{Evolution of the classical action as $\beta$ changes.
(a) $0\le\beta<\pi/\hbar\omega_m$;
(b) $\pi/\hbar\omega_m<\beta<2\pi/\hbar\omega_m$;
(c) $2\pi/\hbar\omega_m<\beta<3\pi/\hbar\omega_m$.}
\label{fig4}
\end{figure}

\begin{figure}
\caption{Partition of control space into $p$-solution regions
$(p=1,3,5,7,\ldots)$.}
\label{fig5}
\end{figure}


\begin{references}

\bibitem[*]{CAC} e-mail: aragao@if.ufrj.br

\bibitem[\dag]{RMC} e-mail: rmc@fis.puc-rio.br

\bibitem{Feynman-Hibbs} R.\ P.\ Feynman and A.\ R.\ Hibbs,
{\it Quantum Mechanics and Path Integrals} (McGraw-Hill, New York, 1965).

\bibitem{Feynman} R.\ P.\ Feynman, {\it Statistical Mechanics}
(Benjamin, Reading, MA, 1972).

\bibitem{Creutz} M.\ Creutz and B.\ Freedman, Ann.\ Phys.\ (NY)
{\bf 132} (1981), 427.

\bibitem{Classical} By classical motion we mean motion satisfying
the Euler-Lagrange equation $M\ddot{x}-V'(x)=0$, which is the
equation of motion for a particle moving in a potential
{\em minus} $V$.

\bibitem{Berry} M.\ Berry, in {\it Physics of Defects}, Les Houches
Session XXXV (1980), eds.\ R.\ Balian et al.\ (North-Holland, Amsterdam,
1981).

\bibitem{Thom} R.\ Thom, {\it Structural Stability and Morphogenesis}
(Benjamin, Reading, MA, 1975); T.\ Poston and I.\ N.\ Stewart,
{\it Catastrophe Theory and its Applications} (Pitman, London, 1978);
E.\ C.\ Zeeman, {\it Catastrophe Theory: Selected Papers 1972-1977}
(Addison-Wesley, Reading, MA, 1977).

\bibitem{Saunders} P.\ T.\ Saunders, {\it An Introduction to
Catastrophe Theory} (Cambridge University Press, Cambridge, 1980).

\bibitem{Peixoto} M.\ M.\ Peixoto and R.\ Thom, C.\ R.\ Acad.\ Sc.\
Paris I, {\bf 303} (1986), 629 and 693; {\bf 307} (1988), 197 (Erratum);
M.\ M.\ Peixoto and A.\ R.\ Silva,
An.\ Acad.\ Bras.\ Ci., {\bf 62} (4) (1990), 321;
M.\ M.\ Peixoto and A.\ R.\ Silva, preprint (1995).

\bibitem{DV} G.\ Dangelmayr and W.\ Veit, Ann.\ Phys.\ (NY)
{\bf 118} (1979), 108.

\bibitem{Weiss} U.\ Weiss, {\it Quantum Dissipative Systems}
(World Scientific, Singapore, 1993).

\end{references}
\end{document}